\documentclass[12pt,a4paper,dvips]{article}
\usepackage{graphicx,epsfig}
\usepackage{hhline}

\usepackage{amsmath,amssymb}
\usepackage{times}
\usepackage[varg]{txfonts}
\DeclareMathAlphabet{\mathbold}{OML}{txr}{b}{it}

\usepackage{array,multirow,dcolumn}
\usepackage[mathlines,displaymath]{lineno}
\usepackage{rotating}
\usepackage[english]{babel}
 
\usepackage[numbers,square,comma,sort&compress]{natbib}
\usepackage{hypernat}
\usepackage{textcomp}
\newcolumntype{.}{D{.}{.}{-1}}
\newcolumntype{-}{D{-}{-}{-1}}

\usepackage[dvipsnames]{color}
\definecolor{rltred}{rgb}{0.75,0,0}
\definecolor{rltgreen}{rgb}{0,0.5,0}
\definecolor{rltblue}{rgb}{0,0,0.5}

\usepackage[hyperindex,bookmarks,bookmarksnumbered,breaklinks,a4paper,unicode]{hyperref}
\hypersetup{%
  pdftitle        = {Measurement of the inclusive ep Scattering Cross Section
 at low Q2 and x at HERA},
  urlcolor        = rltblue,       
  urlbordercolor  = 0 0 0.5,
  filecolor       = rltblue,       
  filebordercolor = 0 0 0.5,
  linkcolor       = rltred,        
  linkbordercolor = 0.75 0 0,
  citecolor       = rltgreen,      
  citebordercolor = 0 0.5 0,
  pagecolor       = rltgreen,      
  pagebordercolor = 0 0.5 0,
  menucolor       = rltgreen,      
  menubordercolor = 0 0.5 0,
  colorlinks    = true,
  pdfauthor     = {H1 Collaboration},
  pdfsubject    = { },
  pdfkeywords   = {High-Energy Physics, Particle Physics, Proton Structure, DIS}
}

\newlength{\dinwidth}
\newlength{\dinmargin}
\newcommand{\ncred}{\mbox{$ \sigma_{r,{\rm NC}}^{\pm}$}}

\newcommand{\ncdd}{\mbox{$\frac{\textstyle {\rm d^2} \sigma^{e^{\pm}p}_{{\rm NC}}}{\textstyle {\rm d}x{\rm d} Q^2}$}}

\begin{document}

\makeatletter \def\NAT@space{} \makeatother

\begin{titlepage}
 
\noindent
DESY 16-064

\vspace*{2.5cm}

\begin{center}
\begin{Large}


{\bfseries Study of HERA {\boldmath{$ep$}} Data at Low {\boldmath{$Q^2$}} 
 and Low {\boldmath{$x_{\rm Bj}$}} and the Need for  
 Higher-Twist Corrections to Standard pQCD Fits}

\vspace*{2cm}

I.~Abt$^a$,
A.M.~Cooper-Sarkar$^b$,
B.~Foster$^{b,c,d}$, 
V.~Myronenko$^d$,
K.~Wichmann$^d$, 
M.~Wing$^e$

\end{Large}
\end{center}

{\protect \hskip 0.cm} $^a$ Max-Planck-Institut f{\"u}r Physik, Werner-Heisenberg-Institut, M{\"u}nchen 80805, Germany 

\vspace*{-.12cm}
{\protect \hskip 0.cm} $^b$ Physics Department, University of Oxford, Oxford OX1 3RH, United Kingdom
 
\vspace*{-.12cm}
{\protect \hskip 0.cm} $^c$ Hamburg University, I. Institute of Experimental Physics, Hamburg 22607, Germany

\vspace*{-.12cm}
{\protect \hskip 0.cm} $^d$ Deutsches Elektronen Synchrotron DESY, Hamburg 22607, Germany

\vspace*{-.12cm}
{\protect \hskip 0.cm} $^e$ Department of Physics and Astronomy, University College London, London WC1E 6BT, United Kingdom

\vspace*{1cm}

\begin{abstract} \noindent

A detailed comparison of HERA data
at low Bjorken-$x$ and low four-momentum-transfer squared, $Q^2$,
with predictions based on 
$\ln{Q^2}$ evolution (DGLAP) in perturbative Quantum Chromo Dynamics
suggests inadequacies of this framework.
The standard DGLAP evolution was augmented by
including an additional higher-twist term in the description of the 
longitudinal structure function, $F_{\rm L}$. 
This additional term, $F_{\rm L}~A_{\rm L}^{\rm HT}/Q^2$, improves  
the description of the reduced cross sections
significantly. The resulting predictions for $F_{\rm L}$ suggest
that further corrections are required for $Q^2$ less than about
2\,GeV$^2$.

\end{abstract}

\vspace*{1.5cm}

\end{titlepage}

\newpage
~~
\newpage

\section{Introduction}
\label{sec:intro}

Analyses of HERA and other DIS data
are generally performed within the
perturbative regime of Quantum Chromo Dynamics (QCD),~\cite{MANDY}
i.e.\ with $Q^2$, the four-momentum-transfer squared, 
sufficiently above 1\,GeV$^2$. 
The HERA data extend towards $Q^2$ and $x_{\rm Bj}$ values, where $x_{\rm Bj}$
is the Bjorken scaling variable,
where the longitudinal structure function, $F_{\rm L}$, becomes significant.
Previous HERA results~\cite{HERAIcombi,HERAPDF20} suggest that QCD 
continues to give a good description of the data down to surprisingly 
low values of $Q^2$.
This gives the possibility of not only establishing the limit below
which QCD no longer describes the data, but also
of investigating modifications to the standard 
Dokshitzer--Gribov--Lipatov--Altarelli--Parisi (DGLAP)~\cite{Gribov:1972ri,Gribov:1972rt,Lipatov:1974qm,Dokshitzer:1977sg,Altarelli:1977zs}
evolution that have been proposed in the literature, for example 
$\ln(1/x)$ resummations, as introduced by 
Balitsky--Fadin--Kuraev--Lipatov~\cite{bfkl}, or "saturation".
The key variable is $x$, the fraction of the proton momentum 
carried by the parton,
which is identical to $x_{\rm Bj}$ in the quark--parton model. 
Saturation is expected to occur
when the density of gluons becomes so large that the standard increase 
in gluon density as $x$ falls is flattened off by 
gluon--gluon interactions and recombination. 
Such effects can be described by non-linear evolution equations
including higher-twist corrections at low $x$, visualised as  
gluon ladders with recombining gluons~\cite{Bartels,Bartels2}.
An earlier analysis of HERA-I data~\cite{caola} had shown some discrepancies with conventional DGLAP evolution 
at low $Q^2$ and low $x$. This is now investigated using the final combination of HERA inclusive cross sections.
 
This combination of
HERA reduced cross sections for neutral current (NC) and charged current (CC) 
$e^{\pm}p$ scattering
measured by the H1 and ZEUS collaborations
was recently published~\cite{HERAPDF20} together with a QCD analysis based
solely on the DGLAP 
formalism, which produced a set of parton distribution
functions (PDFs) called HERAPDF2.0.
In this analysis,
it was noted that the predictions from the PDFs
of HERAPDF2.0 were not able to describe the NC data very well
at low $Q^2$, below $Q^2 \approx 10$\,GeV$^2$, both at NLO and NNLO.
This was confirmed within the framework of the NNPDF global analysis~\cite{rojo}.

The reduced NC deep inelastic $e^{\pm}p$ scattering cross sections 
are given by a linear combination of structure functions which depends on the PDFs. 
At low $Q^2$, where virtual photon exchange is dominant, the reduced cross sections for $e^{\pm}p$
scattering are equal and 
may be expressed in terms of the structure functions $F_2$ and $F_{\rm L}$ as
\begin{eqnarray} \label{ncsi}     
 \ncred =\ncdd \cdot \frac{Q^4 x_{\rm Bj}}{2\pi \alpha^2 Y_+}                                                     
  =            F_2 -\frac{y^2}{Y_+} F_{\rm L}~,
\end{eqnarray}                                                                  
where the fine-structure constant, $\alpha$, 
the photon propagator and a helicity factor are absorbed 
in the definitions of \ncred~and $Y_{\pm}=1 \pm (1-y)^2$.
In particular, the predictions of HERAPDF2.0 were not able to describe
the turn-over of the NC 
reduced cross section at low $x_{\rm Bj}$ and low $Q^2$
due to the contribution from $F_{\rm L}$, which is directly connected 
to the gluon PDF~\cite{CooperSarkar:1988}.

The analysis presented here focuses on a simple $ansatz$ to add
higher-twist terms to the DGLAP-based evolution.
The expectation is that such
terms are important for $F_{\rm L}$,
but not for the structure function $F_2$, because in the case
of $F_2$ longitudinal and transverse contributions cancel~\cite{Bartels}.
New sets of PDFs were extracted at next-to-leading order (NLO)
and next-to-next-to-leading order (NNLO). These PDFs are labelled the HHT PDFs 
and the corresponding analyses are called the HHT analyses, 
for ease of reference.
The predictions from these analyses 
are compared to the reduced HERA cross sections
at  low $Q^2$ and low $x_{\rm Bj}$.
The predictions of HHT and HERAPDF2.0 for $F_{\rm L}$
are compared to measurements published separately by the H1~\cite{H1FL1}
and ZEUS~\cite{ZEUSFL} collaborations.

\section{The HERA Data and HERAPDF2.0}

The HERA data on neutral current and 
charged current $e^+p$ and $e^-p$ inclusive cross 
sections as combined by the H1 and ZEUS collaborations~\cite{HERAPDF20} 
were used as the input 
to the analysis presented here.
Their kinematic range spans six orders of magnitude in
$x_{\rm Bj}$ and $Q^2$, but only four orders of magnitude
are usable for pQCD fits, for which $Q^2_{\rm min}$ must be above
1\,GeV$^2$. The range in $x_{\rm Bj}$ is automatically
reduced when low-$Q^2$ data are excluded, because, at HERA,
low $x_{\rm Bj}$ also implies low $Q^2$. 

The data were previously used to extract
the HERAPDF2.0~\cite{HERAPDF20} set of PDFs. 
While the description of the
data by the predictions of HERAPDF2.0 is quite good, the overall
$\chi^2$/(number of degrees of freedom, ndof)
values of the various fits were around 1.2~\cite{HERAPDF20}.
It was observed that these values could be reduced 
if $Q^2_{\rm min}$, the smallest $Q^2$ of the data used in the fits, was increased from 3.5\,GeV$^2$ to
10\,GeV$^2$. However, this substantially worsened the
predictions for the low-$Q^2$ and low-$x_{\rm Bj}$ regime,
which were already not particularly good either at NLO or NNLO
for the standard fits with $Q^2_{\rm min} = 3.5$\,GeV$^2$.
Neither did NNLO fits show any improvement over NLO ones. 

Most of the HERA data were taken with a centre-of-mass energy, $\sqrt{s}$,
of $318\,$GeV. 
However, NC $e^+p$ data are available also for 
lower $\sqrt{s}$, such that different values
of the inelasticity $y$ are accessed at 
the same $x_{\rm Bj}$ and $Q^2$, since $y=sx/Q^2$. 
This provides direct information on $F_{\rm L}$.
Although results on $F_{\rm L}$  were published separately 
by the H1 and ZEUS collaborations, the data on which the results 
were based were combined and were included in the data set used
for the HHT analysis. 


\section{QCD Analysis Including Higher-Twist Effects}

The introduction of higher-twist terms is one possible
way to extend the DGLAP framework. Higher-twist effects have a $1/Q^{2n}$ dependence and are thus important at low $Q^2$.
Such terms have been introduced by previous authors, but usually in the context of 
higher-twist effects which are important at high $x$~\cite{abm}. In the present paper we investigate 
low-$x$ higher-twist effects since, for the kinematics of HERA, low $Q^2$ is only accessed at low $x_{\rm Bj}$.
Motyka {\it et al.}~\cite{motyka} have also considered higher-twist effects at low $x$ 
but in the context of diffractive data. In the present study we concentrate on inclusive data.
The leading-twist perturbative QCD 
forms of the structure functions $F_2$ and $F_{\rm L}$ 
were augmented by simple twist-4 terms
\begin{eqnarray}
\label{eq:flht}
F_{\rm 2}^{\rm HT} &= F_{2}^{\rm DGLAP} &(1 + A_2^{\rm HT}/Q^2)~,\\
\label{eq:flht2}
F_{\rm L}^{\rm HT} &= F_{\rm L}^{\rm DGLAP} &(1 + A_{\rm L}^{\rm HT}/Q^2)~,
\end{eqnarray}
where $A_2^{\rm HT}$ and $A_{\rm L}^{\rm HT}$ are free parameters  in the fits.

The ZEUSfitter package\,\footnote{The package was recently also used
in a combined electroweak and QCD analysis of HERA data~\cite{ZEUS-EW}.}  
was used for the analysis presented here. The results
were cross-checked with the HERAFitter~\cite{HERAFitter} package.
Except for the addition of the higher-twist term,
the fits called HHT were set up exactly as the HERAPDF2.0 fits. In particular the 
heavy-flavour scheme used was the
RTOPT scheme~\cite{Thorne:1997ga,Thorne:2006qt,Thorne:RTopt} and the minimum value of $Q^2$ for data entering the fit was $Q^2_{\rm min}=3.5$\,GeV$^2$.
The value of $Q^2_{\rm min}=3.5$\,GeV$^2$ was chosen
as it was assumed that non-perturbative effects would only appear at $Q^2$ below this value.
Other groups
work with even lower $Q^2_{\rm min}$, for example
MSTW/MMHT~\cite{Martin:2009iq,MMHT2014}
use $Q^2_{\rm min}=2.0$\,GeV$^2$. 
A higher-twist term as introduced in Eq.~\ref{eq:flht2} was tested by MMHT 
and found to improve the
$\chi^2$ values of their fits~\cite{MMHT-HT-16}
to HERA and other data. In the present paper the effects of such a higher-twist term on the 
predictions for $F_2$ and \ncred~ are explored in more detail, using an analysis focussing on HERA data.

The  PDFs for HHT were parameterised
at the starting scale $\mu^2_{{\rm f}_0}=1.9$\,GeV$^2$.
The gluon PDF, directly connected to $F_{\rm L}^{\rm DGLAP}$, 
was parameterised as
\begin{equation}
\label{eq:xgpar}
xg(x) =   A_g x^{B_g} (1-x)^{C_g} - A_g' x^{B_g'} (1-x)^{C_g'} ~ ,  \\
\end{equation}
where $A_g, B_g, C_g$ and $A_g',B_g'$ are free parameters 
and ${C_g'}$ was set to $25$~\cite{Martin:2009iq}.
The $A_g'$ was added to make the parametersiation more
flexible at low $x$. 
It could lead to a negative gluon density at low $x$, 
even at scales above $\mu^2_{{\rm f}_0}$. However, this 
was neither observed for HERAPDF2.0 nor for
the analysis presented here.


The HHT fits were performed at NNLO and NLO,
including the higher-twist term for $F_2$
only, $F_{\rm L}$ only and both  $F_2$ and $F_{\rm L}$.
The uncertainties from the fits are taken
as experimental uncertainties and are
the only uncertainties considered throughout
the paper.
The introduction of  $A_{\rm L}^{\rm HT}$ was found to reduce
the $\chi^2$/ndof of the fit significantly, both at NLO and at NNLO.
However, adding $A_2^{\rm HT}$ had no significant effect.
For the NNLO fit, it only reduced the  $\chi^2$/ndof  from 1363/1131 to 
1357/1130  and the corresponding value of $A_2^{\rm HT}$ was  
consistent with zero, i.e.\ $A_2^{\rm HT} =0.12\pm 0.07$\,GeV$^2$. 
Similar values for $A_2^{\rm HT}$ were obtained when $A_2^{\rm HT}$ and
$A_{\rm L}^{\rm HT}$ were included simultaneously.
Therefore, all HHT fits presented in this paper 
include only  the $A_{\rm L}^{\rm HT}$ term.
This agrees with  predictions~\cite{Bartels} that
higher-twist terms would be observable in $F_{\rm L}$ but not in
$F_2$ because the contributions from longitudinally and
transversely polarised photons would cancel for $F_2$.

\begin{figure}[tbp]
\vspace{-0.5cm} 
\centerline{
  \epsfig{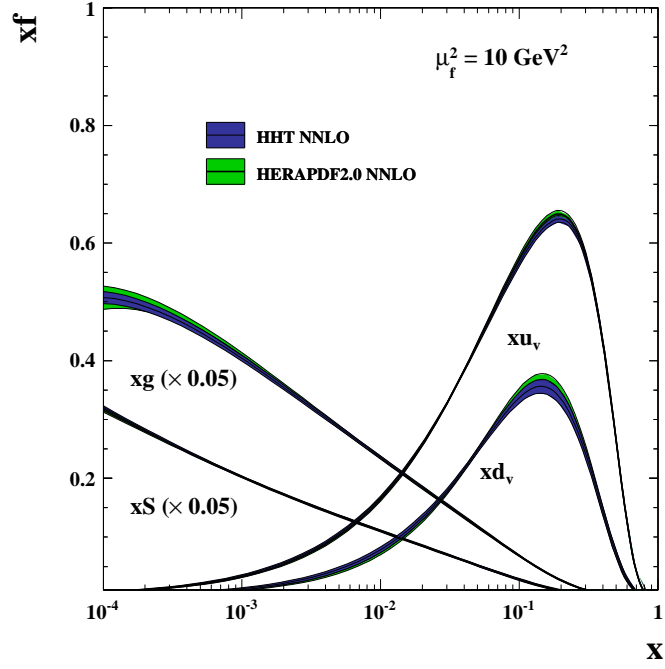}
}
\centerline{
  \epsfig{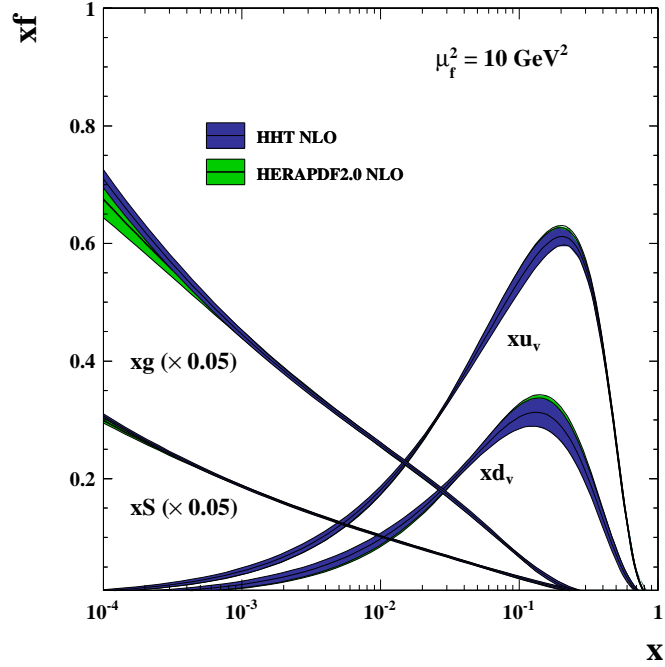}
}
\caption { 
      The HHT parton distribution functions,  
      $xu_v$, $xd_v$, $xS$
      and $xg$, at the scale
      $\mu_{\rm f}^2 = 10$\,GeV$^2$ compared to the PDFs from 
      HERAPDF2.0 at NNLO~(top) and NLO~(bottom). 
      The gluon and sea distributions are scaled down by a factor $20$.
      The bands represent the experimental, i.e.\ fit, uncertainties.
}
\label{fig:summary}
\end{figure}

The HHT PDFs, $xd_v$ and $xu_v$ for the valence quarks and
$xS$ for the sea quarks together with $xg$, 
are shown in Fig.~\ref{fig:summary}.
The PDFs of HHT are very similar to the PDFs of HERAPDF2.0,
even though
the values of $A_{\rm L}^{\rm HT}$ extracted are quite high: 
$A_{\rm L}^{\rm HT}=5.5\pm0.6$~GeV$^2$ from the NNLO  and
$A_{\rm L}^{\rm HT}=4.2\pm0.7$~GeV$^2$ from the NLO fit.
The PDFs of HHT remain very similar to those of HERAPDF2.0 when they 
are evolved in $Q^2$ up to the scale of the LHC, across the kinematic reach of $x_{\rm Bj}$
of the ATLAS, CMS and LHCb experiments.
Thus the need for higher-twist terms at low $Q^2$ has no effect on LHC physics.

The $\chi^2$/ndof 
for HHT NNLO is $1316/1130$ 
and for HHT NLO $1329/1130$.
This may be compared to the HERAPDF2.0 $\chi^2$/ndof values
of $1363/1131$ for the  NNLO and
$1356/1131$ for the NLO fit.  
This represents an improvement of 
$\Delta\chi^2= -27$
for NLO and an even more significant $\Delta\chi^2 = -47$ at NNLO.
Table~\ref{tab:chitab} details 
the main contributions to this reduction of $\chi^2$.
The HHT fit at NNLO has a lower $\chi^2$ than the fit at NLO.
This is a reversal of the situation for HERAPDF2.0.
Table~\ref{tab:chitab} also lists the partial $\chi^2$/ndp values
for the high-precision NC $e^+p$ data with $\sqrt{s}=318$\,GeV$^2$.
In addition, the $\chi^2$/ndp values for the data points below
$Q^2_{\rm min}=3.5$\,GeV$^2$ down to 2.0\,GeV$^2$ are listed.
These $\chi^2$ values provide an
evaluation of the quality of the predictions below $Q^2_{\rm min}$
and quantify that the extrapolation of HHT NNLO 
describes these data better than the extrapolation of HERAPDF2.0,
while the description at NLO does not improve.

The positive higher-twist terms preferred by the HHT fits imply that $F_{\rm L}$ is larger than determined in the HERAPDF2.0 
fits. Since the structure function $F_{\rm L}$ is directly related to the gluon
distribution at low  $x$, at $\mathcal{O}(\alpha_s)$, it might be expected that a larger $F_{\rm L}$ implies at larger low-$x$ gluon.
However, this ignores the role of higher-order matrix elements. In fact, the NNLO gluon distribution 
exhibits a turn-over at 
low $x$ and $Q^2$.  This comes from the 
substantial $A_g'$ term which the HHT NNLO fit requires even in the presence of the 
large higher-twist term. To investigate this a gluon parameterisation of the form $xg(x) =   A_g x^{B_g} (1-x)^{C_g}(1+D_gx)$
was also tested at both NLO and NNLO. This form is called the alternative gluon or AG form of the parametrisation and
it ensures that the gluon distribution is always 
positive definite for $Q^2 \ge \mu^2_{{\rm f}_0}$.
The AG fits and the fits using the form of Eq.~\ref{eq:xgpar} are very similar at NLO.
In contrast, the AG parameterisation 
at NNLO results in much higher $\chi^2$/ndof values,
1389/1133 for HERAPDF2.0 and 1350/1132 for HHT. 
At NNLO the data favour a strong gluon turn-over 
whereas AG, by construction,
does not allow this. The AG parameterisation
is clearly not suited for fits at NNLO.

\begin{figure}[tbp]
\centerline{
\epsfig{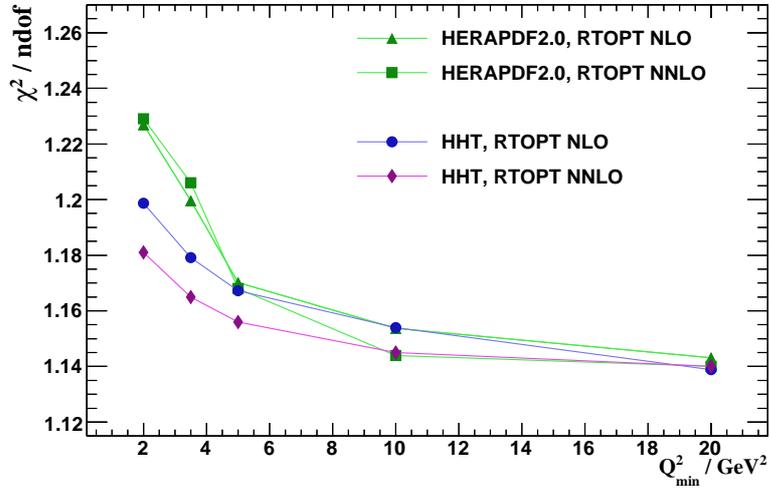}}
\caption { 
The $\chi^2$/ndof versus $Q^2_{\rm min}$ for HHT and HERAPDF2.0 fits
at NNLO and NLO.
}
\label{fig:chiscan}
\end{figure}

The validity of the assumption that 
perturbation theory is applicable
in the kinematic regime of the fits
is tested by the dependence
of the quality of the fits, as
represented by $\chi^2$/ndof, on the value of $Q^2_{\rm min}$.
The value of $\chi^2$/ndof should ideally not depend strongly
on $Q^2_{\rm min}$.
The dependence of $\chi^2/$ndof
on $Q^2_{\rm min}$ 
for HHT and HERAPDF2.0 
is shown in Fig.~\ref{fig:chiscan} for both NNLO and NLO.
The values drop steadily until 
$Q^2_{\rm min} \approx 10~$GeV$^2$, when the $\chi^2$/ndof 
becomes similar for HHT and HERAPDF2.0.
The effect of the higher-twist term is, as expected, confined
to the low-$Q^2$ region.
The HHT fits show a slower rise in $\chi^2$ as $Q^2_{\rm min}$ is reduced.

The fits with $Q^2_{\rm min}=2.0$\,GeV$^2$
close to the starting scale $\mu^2_{{\rm f}_0}=1.9$\,GeV$^2$
were studied in more detail.  
The relevant $\chi^2$ values are listed in Table~\ref{tab:chitab2}.
The PDF and especially the higher-twist parameters 
of HHT NNLO do not change much when
$Q^2_{\rm min}$ is lowered from 3.5\,GeV$^2$ to 2.0\,GeV$^2$. 
The partial $\chi^2$/ndp for the NC $e^+p$ data with
$\sqrt{s}=318$\,GeV increases from 1.12 to 1.14, but the
partial $\chi^2$/ndp drops from 1.28 to 1.04
for the 25 points in the range $2.0 \le Q^2 < 3.5$\,GeV$^2$. 

Refitting with lower $Q^2_{\rm min}$ has a stronger 
effect at NLO than at NNLO, but again, the higher-twist term 
is basically unchanged. 
The results at NLO are, as before, not strongly dependent on 
the details of the gluon distribution. This can be seen when refitting 
with HHT NLO AG, which yields almost the same result as HHT NLO.

\section{Heavy-Flavour Schemes}

\begin{figure}[tbp]
\centerline{
\epsfig{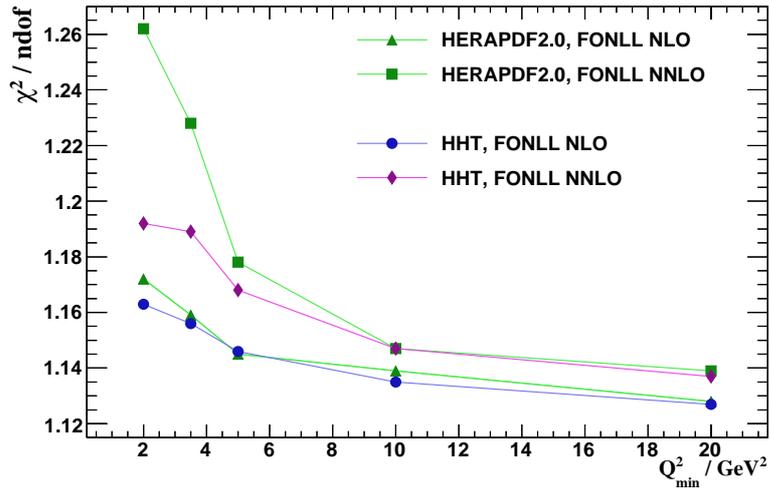}}
\caption { 
The $\chi^2$/ndof versus $Q^2_{\rm min}$ for HHT and HERAPDF2.0 fits
at NNLO and NLO using the FONLL heavy-flavour scheme instead of the
default RTOPT scheme.
}
\label{fig:chiscan-FONL}
\end{figure}

The influence of the heavy-flavour scheme was already discussed in
the context of HERAPDF2.0~\cite{HERAPDF20}. 
To study the effect on this analysis, 
the HERAFitter~\cite{HERAFitter} package was used 
to replace the default RTOPT scheme with the
fixed-order plus next-to-leading logarithms (FONLL) scheme~\cite{Cacciari:1998it,Forte:2010ta}. 
The resulting dependence of $\chi^2$ on $Q^2_{\rm min}$ 
is shown in Fig.~\ref{fig:chiscan-FONL}, together with the values
from the standard fits.

In the FONLL scheme, the HHT NNLO fit has a substantially  
improved  $\chi^2$/ndof for low $Q^2_{\rm min}$  compared 
to HERAPDF2.0, just as for the standard HHT NNLO fit with RTOPT.  
The value of the higher-twist parameter $A_{\rm L}^{\rm HT}=6.0\pm 0.7$\,GeV$^2$ 
is also similar. 
However, the HHT NLO FONLL fit has only a marginally 
improved $\chi^2$/ndof for low $Q^2$ as compared to 
HERAPDF2.0 and a small value of $A_{\rm L}^{\rm HT}$, 
i.e.\ $A_{\rm L}^{\rm HT}=1.2 \pm 0.6$\,GeV$^2$. This is probably 
associated with the order of $\alpha_s$ at which 
$F_{\rm L}$ is evaluated in these different heavy-flavour schemes. 
RTOPT at NLO calculates $F_{\rm L}$  to $\mathcal{O}(\alpha_s^2)$ and RTOPT at NNLO calculates $F_{\rm L}$  to $\mathcal{O}(\alpha_s^3)$. FONLL at NLO calculates $F_{\rm L}$  to $\mathcal{O}(\alpha_s)$ and FONLL at NNLO calculates $F_{\rm L}$  
to $\mathcal{O}(\alpha_s^2)$.
Only calculating $F_{\rm L}$ to $\mathcal{O}(\alpha_s)$ results 
in  a relatively large $F_{\rm L}$, which can reduce the need for
a higher-twist term.
However, as soon as $F_{\rm L}$ is calculated 
to $\mathcal{O}(\alpha_s^2)$ or higher,  
a higher-twist term is required.
The best fit achieved for HHT NNLO is with the RTOPT scheme.

\section{Reduced Cross Sections}

\begin{figure}[tbp]
\vspace{-0.3cm} 
\centerline{
\epsfig{figure=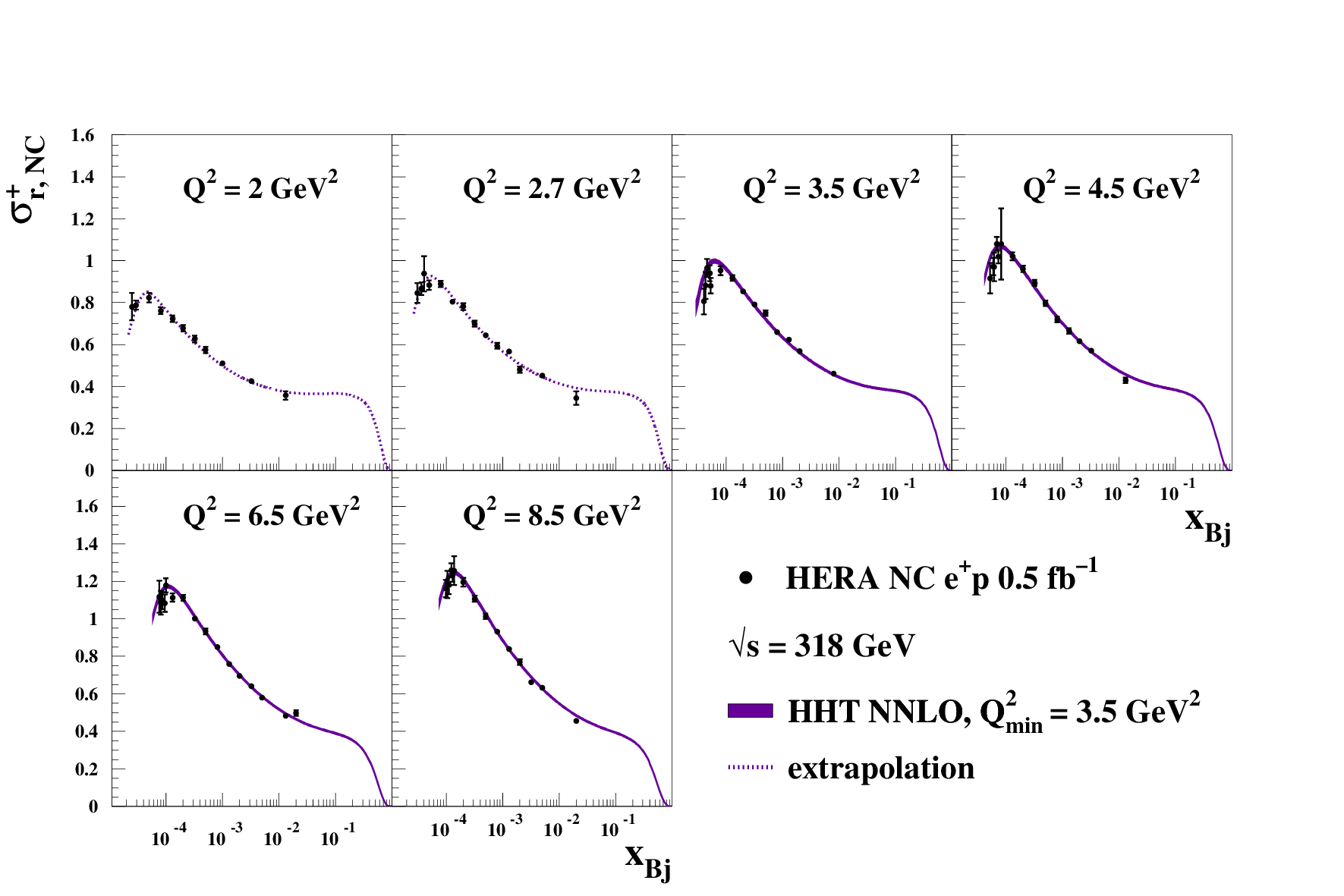 ,width=0.9\textwidth}}
\centerline{
\epsfig{figure=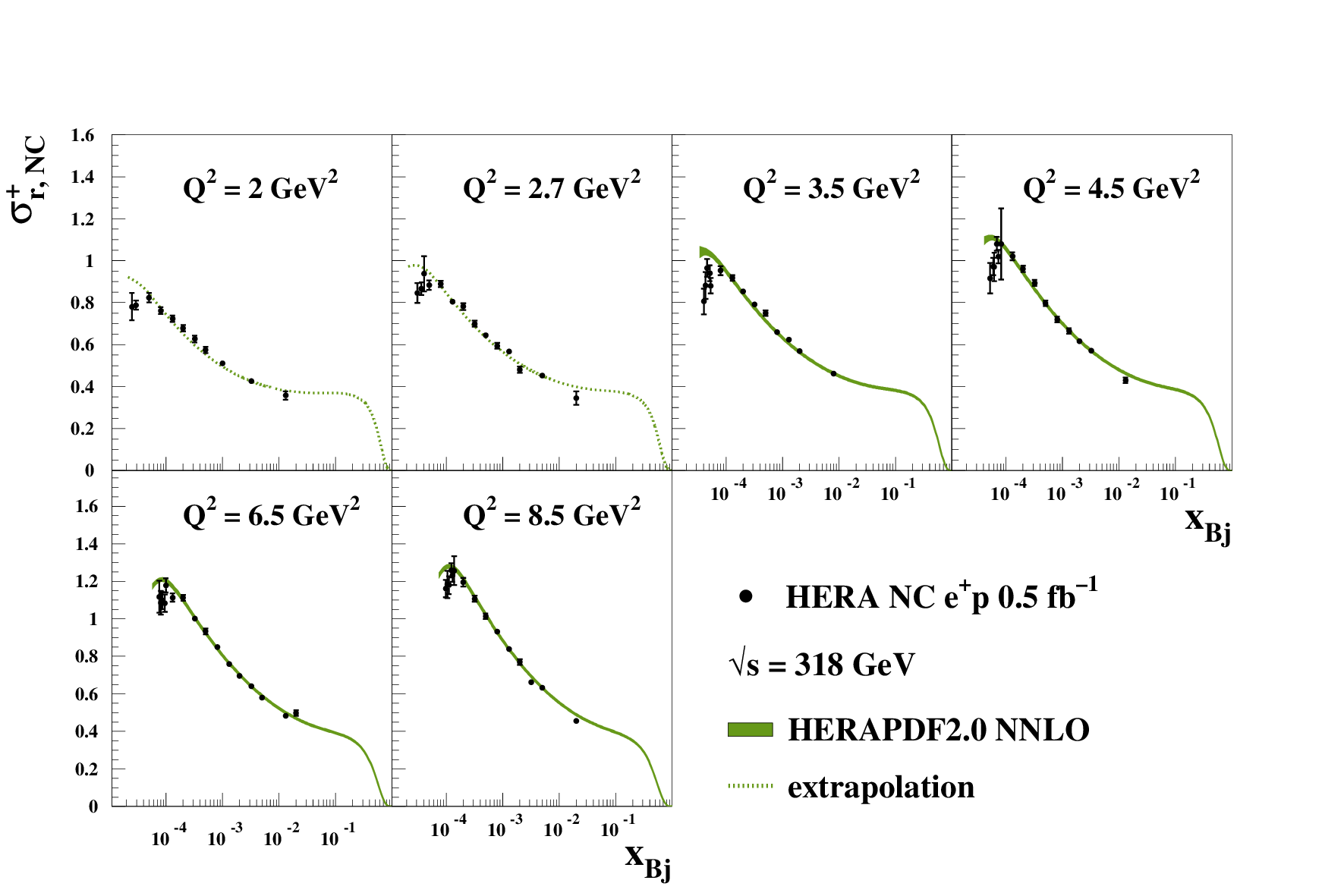,width=0.9\textwidth}}
\vspace{0.5cm}
\caption {The predictions of HHT NNLO (top) and HERAPDF2.0 NNLO 
          (bottom),
          both with $Q^2_{\rm min} = 3.5$\,GeV$^2$, 
          compared to the HERA measurements of $\sigma_{r}$. 
          The bands represent the experimental, i.e.\ fit, uncertainties.
          Extrapolations are indicated as dotted lines.
}
\label{fig:NNLO:HHT-HPDF}
\end{figure}

A comparison  
of the predictions
of HHT and HERAPDF2.0 with $Q^2_{\rm min} = 3.5$\,GeV$^2$
to the measured reduced NC $e^+p$ cross sections is 
shown at NNLO in Fig.~\ref{fig:NNLO:HHT-HPDF}  
and at NLO in Fig.~\ref{fig:NLO:HHT-HPDF}.
In all cases, the predictions are
extrapolated down to $Q^2 = 2.0$\,GeV$^2$;
HHT clearly describes this low-$Q^2$, low-$x_{\rm Bj}$ data better.
This was already indicated by the
$\chi^2$/ndof values in Table~\ref{tab:chitab}, 
where the $\chi^2$/ndp 
for the data points with $2.0 \le Q^2 < 3.5$\,GeV$^2$ are 
listed separately.
The HHT NNLO predictions are clearly preferred as
they describe the turn-over of the data towards low $x_{\rm Bj}$ quite well.
This turn-over region at low $x_{\rm Bj}$ is not well described
by the predictions from HERAPDF2.0.

\begin{figure}[tbp]
\vspace{-0.3cm} 
\centerline{
\epsfig{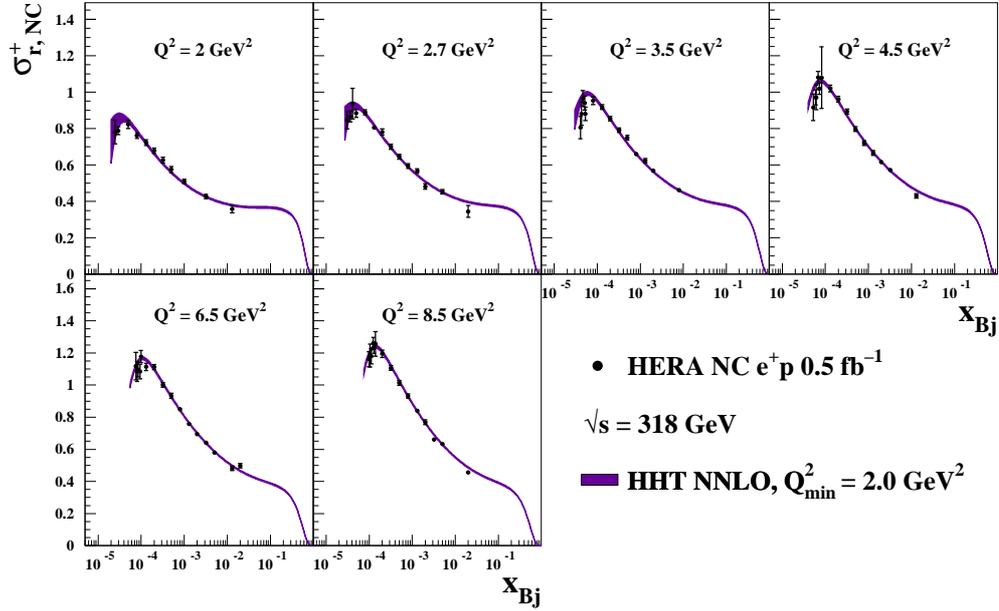}}
\centerline{
\epsfig{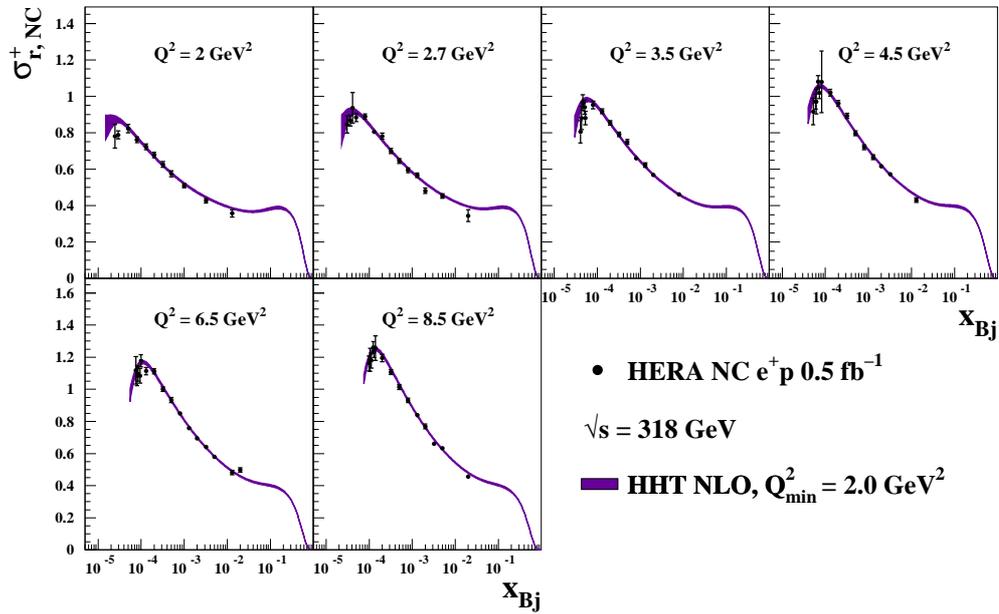}}
\vspace{0.5cm}
\caption {The predictions of HHT NNLO (top) and HHT NLO (bottom) 
          with $Q^2_{\rm min}=2.0$\,GeV$^2$ compared to
          the HERA measurements of $\sigma_r$. 
          The bands represent the experimental, i.e.\ fit, uncertainties.
}
\label{fig:sig_lowq2}
\end{figure}

The predictions of the HHT NNLO and HHT NLO with 
$Q^2_{\rm min} = 2.0$\,GeV$^2$
are shown in Fig.~\ref{fig:sig_lowq2}.
The data are  well described at NNLO, even better than for the
standard HHT NNLO with $Q^2_{\rm min} = 3.5$\,GeV$^2$.
The effect of the lower $Q^2_{\rm min}$ is stronger
at NLO, where the turn-over is better described.

\begin{figure}[tbp]
\vspace{-0.3cm} 
\centerline{
\epsfig{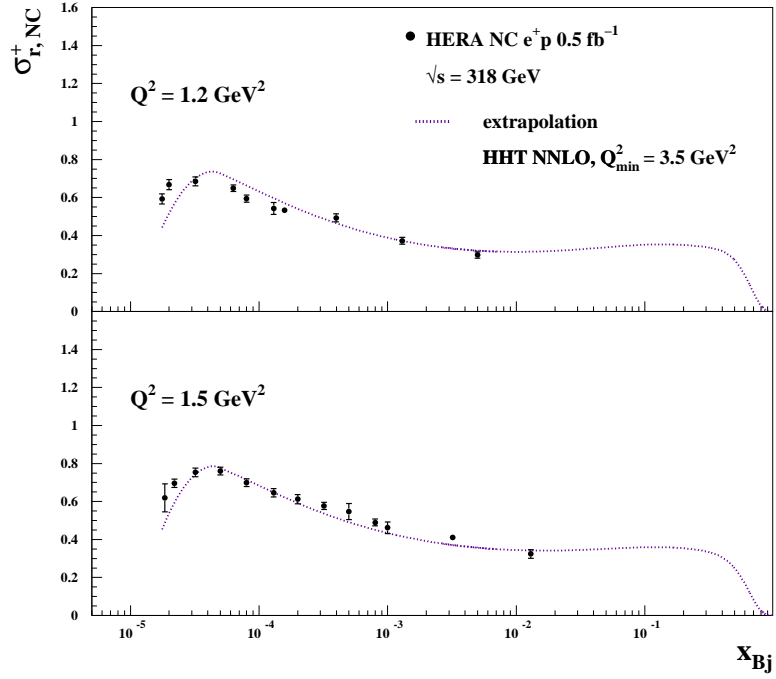}}
\centerline{
\epsfig{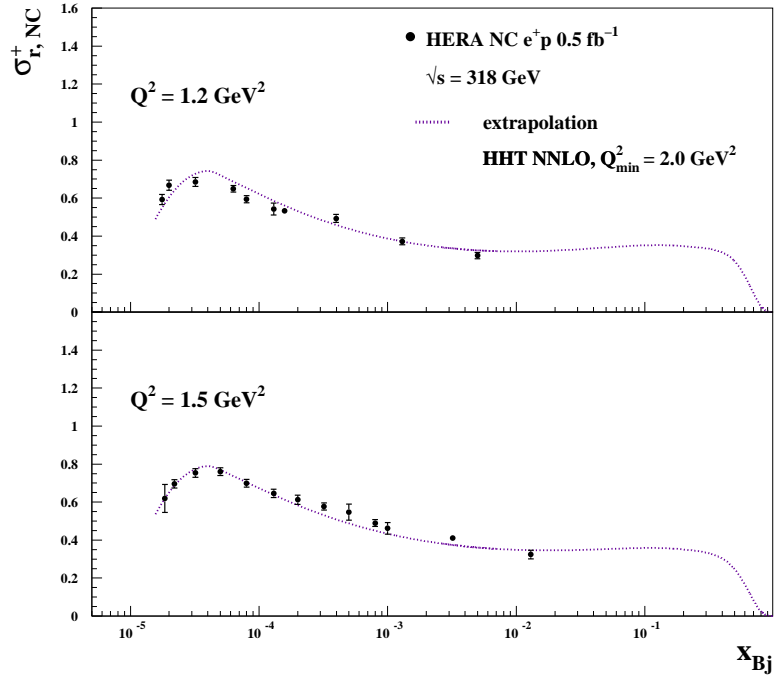}}
\vspace{0.5cm}
\caption {The extrapolated predictions of HHT NNLO  with
          with $Q^2_{\rm min}=3.5$\,GeV$^2$ (top)
          and  with $Q^2_{\rm min}=2.0$\,GeV$^2$ (bottom)
          compared to the HERA NC $e^+p$ measurements of $\sigma_r$
          at $Q^2$ of 1.2 and 1.5\,GeV$^2$. 
}
\label{fig:sig_lowq2:1.2}
\end{figure}

The HHT NNLO  predictions even describe the data down to 
$Q^2 = 1.2$\,GeV$^2$ 
quite well, as can be seen in Fig.~\ref{fig:sig_lowq2:1.2}.
This is especially true for HHT NNLO with $Q^2_{\rm min} = 2.0$\,GeV$^2$.
At $Q^2 = 1.5$\,GeV$^2$, the turn-over is very well described.
At $Q^2 = 1.2$\,GeV$^2$, the predicted turn-over is
somewhat shifted towards higher $x_{\rm Bj}$.
However, it is remarkable how well these data below the starting scale
of the evolution are described,
illustrating once again the 
apparent ability of a perturbative QCD $ansatz$
to describe the data to surprisingly low $Q^2$.

\section{The Structure Functions $F_2$ and $F_{\rm L}$} 

\begin{figure}[tbp]
\vspace{-0.3cm} 
\centerline{
\epsfig{figure=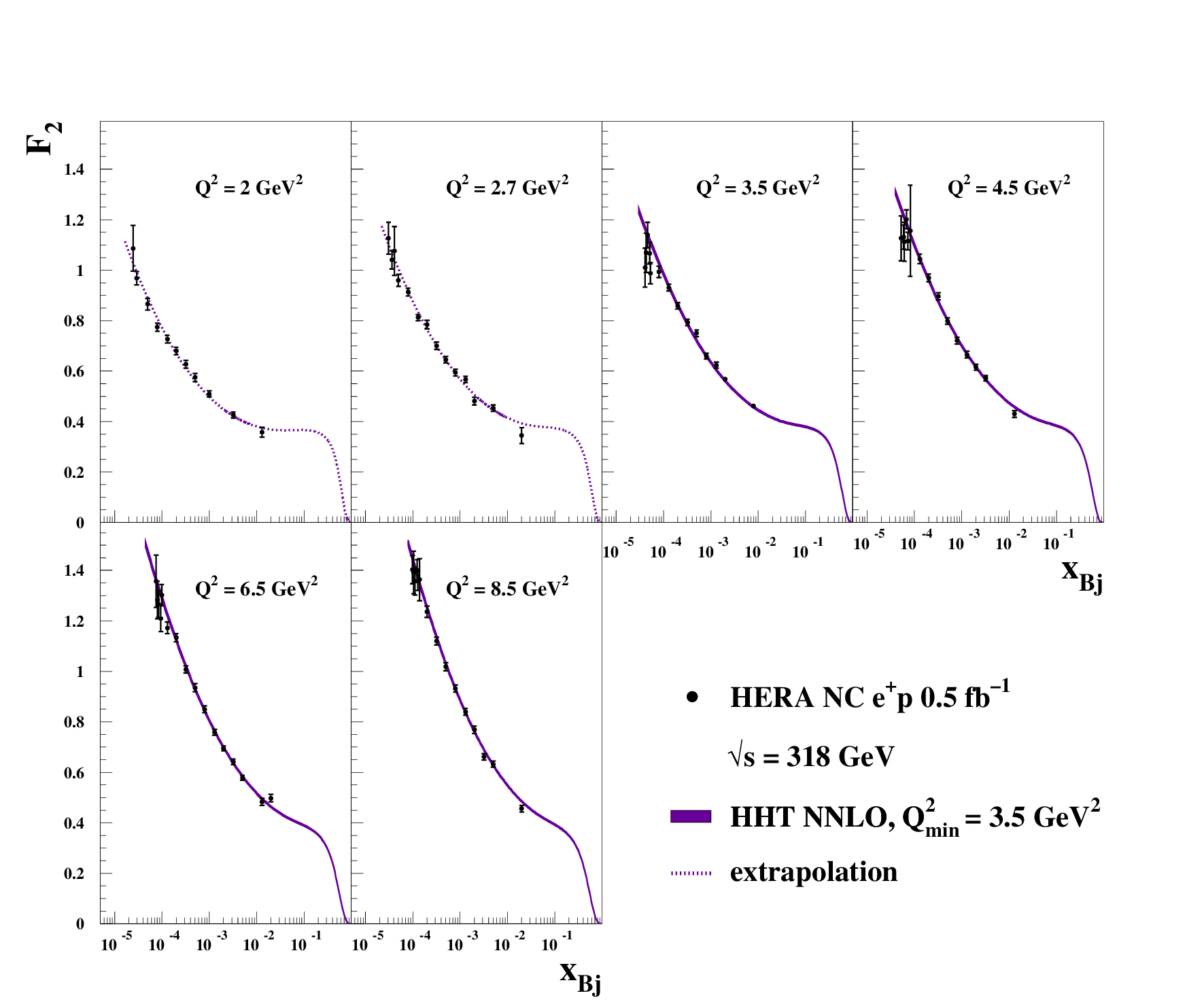 ,width=0.7\textwidth}}
\centerline{
\epsfig{figure=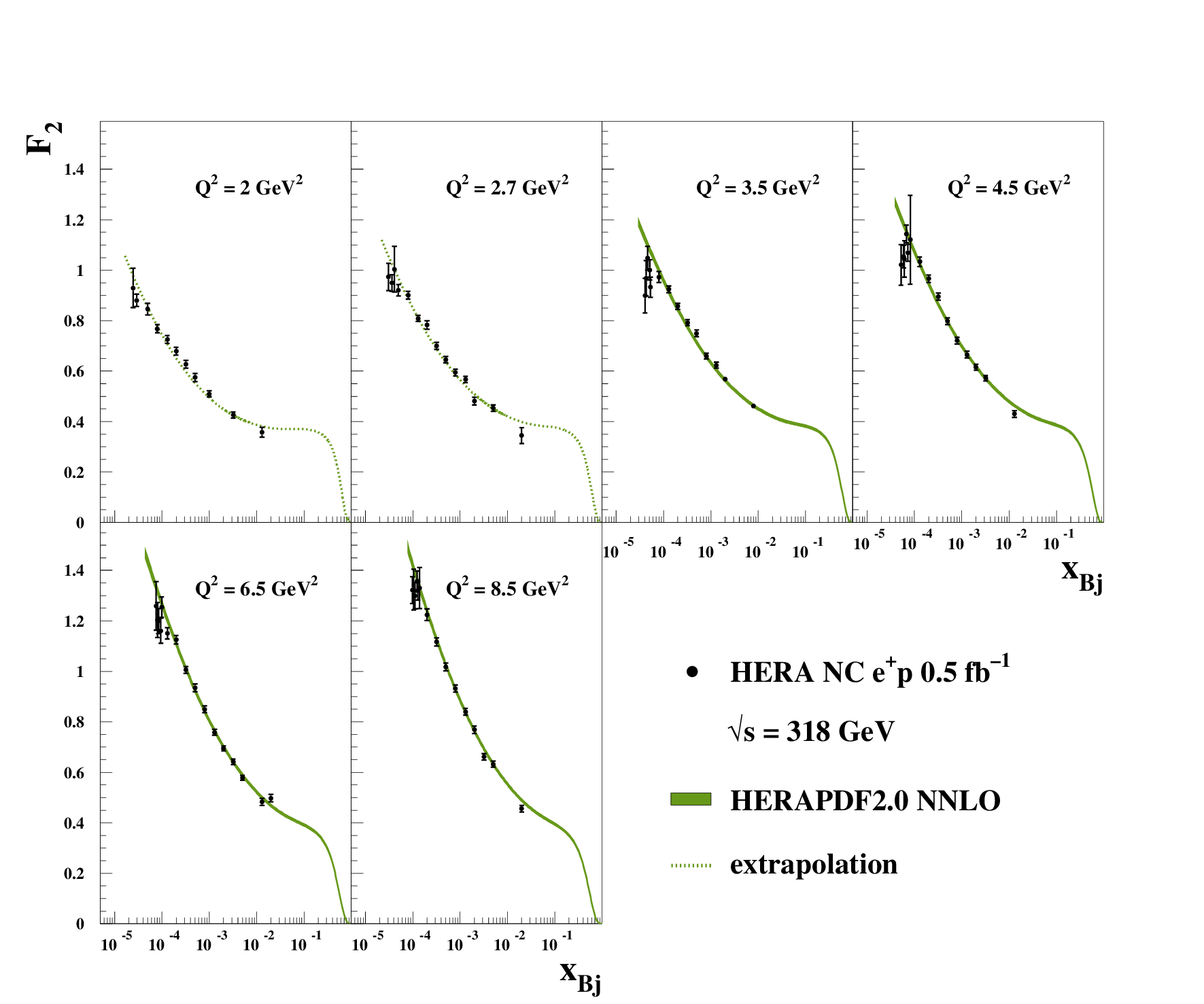  ,width=0.7\textwidth}}
\vspace{0.5cm}
\caption {The predictions of HHT NNLO (top) and HERAPDF2.0 NNLO (bottom),
         both with $Q^2_{\rm min} = 3.5$\,GeV$^2$, 
         compared to extracted values $F_2^{\rm extracted}$. 
         The bands represent the experimental, i.e.\ fit, uncertainties.
         Extrapolations are indicated as dotted lines.
}
\label{fig:f2:NNLO:HHT-HPDF}
\end{figure}

Values of the structure function $F_2$ are extracted from the
data as 
\begin{equation}\label{eq:f2extr}
 F_2^{{\rm extracted}} = F_2^{{\rm predicted}} \frac{\sigma_r^{{\rm measured}}}
                                        {\sigma_r^{{\rm predicted}}} ~~.
\end{equation}
The values of $F_2^{{\rm extracted}}$ 
together with  $F_2^{{\rm predicted}}$ are shown
in Figs.~\ref{fig:f2:NNLO:HHT-HPDF} and~\ref{fig:f2:NLO:HHT-HPDF} 
for HHT and HERAPDF2.0 at NNLO and NLO, respectively.
At NNLO, the HHT predictions and extractions agree well down
to $Q^2=2.0$\,GeV$^2$. 
Since $A_{\rm L}^{\rm HT}$ is substantial, the predictions
from HHT for $F_{\rm L}$ are larger than  
from HERAPDF2.0 at low $Q^2$.
Since $\sigma_r = F_2 - F_{\rm L}~y^2/Y_+$, see Eq.~\ref{ncsi},
this results also in larger predictions for $F_2$ and in larger values
of $F_2^{{\rm extracted}}$. The agreement between prediction and
extraction is better for HHT.  
This confirms that the $F_{\rm L}$ from HERAPDF2.0 
is not large enough. 
The predicted and the extracted values also
agree better for HHT at NLO, but the NLO fit is not as good as the
NNLO fit below
around $Q^2=4.5$\,GeV$^2$.

\begin{figure}[tbp]
\vspace{-0.3cm} 
\centerline{
\epsfig{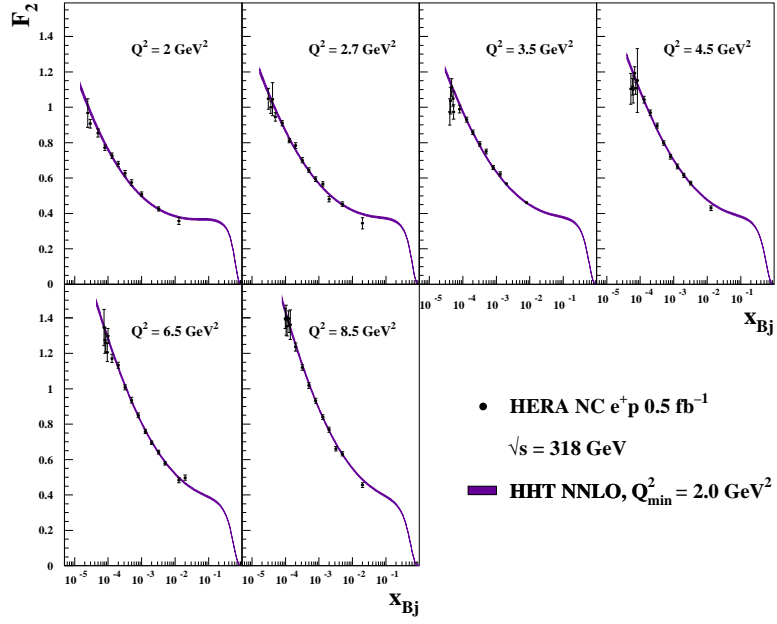}}
\centerline{
\epsfig{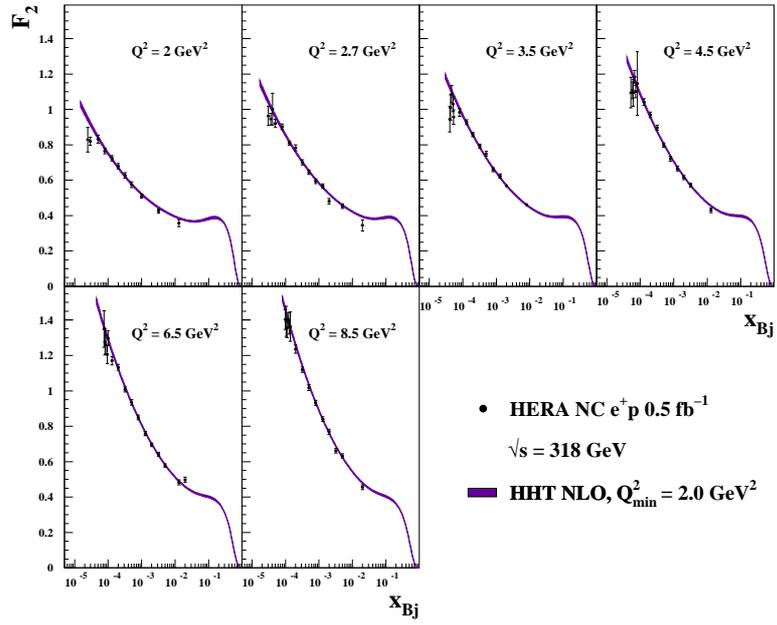}}
\vspace{0.5cm}
\caption {The predictions of HHT NNLO (top) and HHT NLO (bottom) 
         with $Q^2_{\rm min} = 2.0$\,GeV$^2$
         compared to extracted values $F_2^{\rm extracted}$. 
         The bands represent the experimental, i.e.\ fit, uncertainties.
}
\label{fig:f2htq2gt2}
\end{figure}

In Fig.~\ref{fig:f2htq2gt2}, $F_2^{{\rm predicted}}$ 
and $F_2^{{\rm extracted}}$ are shown for HHT NNLO and NLO  
with $Q^2_{\rm min}=2.0$\,GeV$^2$.
The situation for the NNLO fit looks very similar to the fit
with  $Q^2_{\rm min}=3.5$\,GeV$^2$.
The description of the data 
by the predictions of the  NLO fit is improved
at low $x_{\rm Bj}$ and low $Q^2$.
However, $F_2^{{\rm extracted}}$ still shows a tendency to turn-over.
This confirms the findings of the comparisons 
with the reduced cross-section data
that HHT NNLO is better suited to describe the data
than HHT NLO.

\begin{figure}[tbp]
\vspace{-0.3cm} 
\centerline{
\epsfig{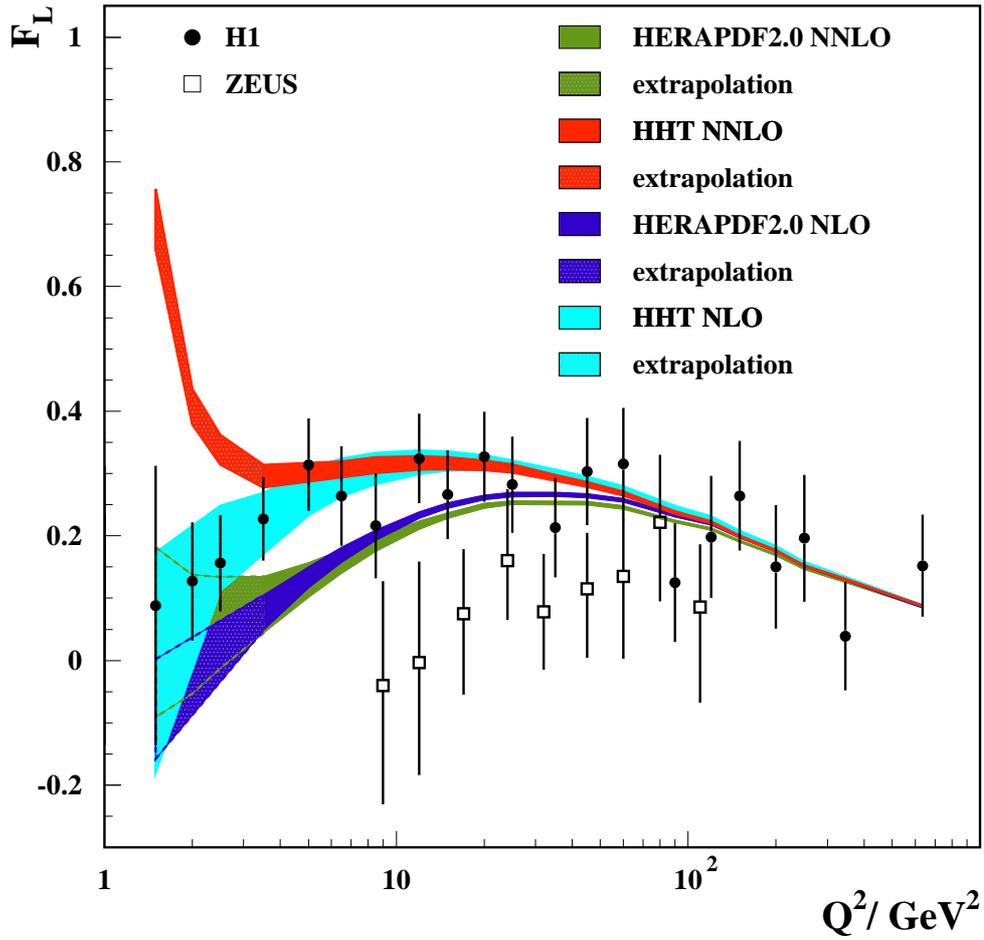}}
\vspace{0.5cm}
\caption {The predictions for $F_{\rm L}$ from HHT and HERAPDF2.0,
          both with $Q^2_{\rm min} = 3.5$\,GeV$^2$, 
          compared to the separate direct measurements published by the 
          H1 and ZEUS collaborations.
          The bands represent the experimental, i.e.\ fit, uncertainties 
          on the predictions. Hatched bands represent extrapolations.
}
\label{fig:flht}
\end{figure}

The H1 and ZEUS collaborations published separate 
measurements of $F_{\rm L}$~\cite{H1FL1,ZEUSFL}, 
using data with lowered $\sqrt{s}$ which provided cross sections
at different $y$ values for identical $x_{\rm Bj}$ and $Q^2$.
The predictions of HHT and HERAPDF2.0 
with $Q^2_{\rm min} = 3.5\,$GeV$^2$ for $F_{\rm L}$ at NNLO and NLO
are compared to these measurements in Fig.~\ref{fig:flht}.
For $Q^2 > Q^2_{\rm min}$, the shapes of all predicted curves
are similar but the predictions of HHT are significantly
higher than those from HERAPDF2.0 for $Q^2$ below 50\,GeV$^2$. 
Even though the statistical accuracy of
the data is limited, the data
mildly favour HHT over HERAPDF2.0 in this regime.

In Fig.~\ref{fig:flht}, extrapolated $F_{\rm L}$ predictions
are shown below  $Q^2_{\rm min}=3.5\,$GeV$^2$ and even 
below the starting scale
$\mu^2_{{\rm f}_0}=1.9$\,GeV$^2$.  
These predictions have large uncertainties and the accuracy
of the data is limited, but
it is clear that the upturn of $F_{\rm L}$ predicted by 
HHT NNLO is not favoured by the data.
This disagreement on  $F_{\rm L}$ is in contrast to          
the fact that the predictions of HHT NNLO describe the
very precise NC $e^+p$ cross sections down to $Q^2=1.2\,$GeV$^2$
remarkably well, see Fig.~\ref{fig:sig_lowq2:1.2}.
Although the higher-twist term is expected to be important for
$F_{\rm L}$~\cite{Bartels}, the very large increase of the 
predicted $F_{\rm L}$ suggests that some other effect is being absorbed in $F_{\rm L}$ 
in the simple $ansatz$  used in the current analysis. Since HERA kinematics couples 
low $Q^2$ to low $x$ it 
could be that $\ln(1/x)$ resummation has a role to play here.

Interestingly, HERAPDF2.0 NNLO also predicts a slight upturn
of $F_{\rm L}$ at $Q^2$ below $Q^2_{\rm min}$. This suggests that the upturn in both the HHT 
and HERAPDF2.0 NNLO analyses
is connected to
the NNLO coefficient functions, which are large and positive.
Similar effects were observed previously~\cite{RT:FL:06} 
for predictions from both pure DGLAP analyses and 
those including higher-twist terms.

\section{Saturation}

The operator-product expansion beyond leading 
twist has diagrams in which two, three or four gluons 
may be exchanged in the t-channel such that  
these gluons may be viewed as recombining. This recombination 
could lead to gluon saturation~\cite{Bartels2}.
The $A^{\rm HT}_{\rm L}/Q^2$-term used in the analysis 
presented here corresponds to twist-4.
Another approach to describe saturation is the 
colour-dipole picture, which is formulated in the proton
rest frame where the incoming photon develops structure over a
coherence length proportional to $1/Q^2$ and $1/x_{\rm Bj}$.
Recently, fits to the HERA data were presented~\cite{Allen}, which
indicate that saturation effects should set in at
latest at $x_{\rm Bj} > 10^{-9}$, 
but possibly earlier.
The data presented here reach down to $x_{\rm Bj} \approx 10^{-5}$. 
It is therefore interesting to see
if there is any hint of saturation effects becoming important 
already in these HERA data.

Phenomenological models of saturation have been treated in the colour-dipole picture.
A successful dipole model using the non-linear running-coupling Balitsky-Kovchegov equation~\cite{balitsky,kovchegov}
 has been developed by Albacete {\it et al.}~\cite{albacete}. However, fits in such a scheme are beyond the
 scope of the present paper. Instead the HERA data are here compared with the predictions of a simple
dipole model
of saturation~\cite{gb} by Golec-Biernat and W{\"u}sthoff (GBW),
in which the onset of saturation is characterised
as the transition from a ``soft'' to a ``hard'' scattering regime.
This occurs along a ``critical line'' in the $x_{\rm Bj}, Q^2$ plane.
Fits to early HERA data with low $Q^2$ and low $x_{\rm Bj}$
indicated that
the criticial line would be around
$x_{\rm Bj} = 10^{-4}$ at $Q^2=1$\,GeV$^2$ and
$x_{\rm Bj} = 10^{-5}$ at $Q^2=2$\,GeV$^2$~\cite{gb}.
These very low-$Q^2$ and low-$x_{\rm Bj}$ data are mostly not included in the
present HHT analysis. This analysis is based on the DGLAP formalism 
which is not expected to
work for $Q^2$ as low as 1\,GeV$^2$. 
The necessary $Q^2_{\rm min}$ cut limits the range 
of the fitted data in $x_{\rm Bj}$ such that the data used here
just touch the predicted critical line.

\begin{figure}[tbp]
\vspace{-0.3cm} 
\centerline{
\epsfig{figure=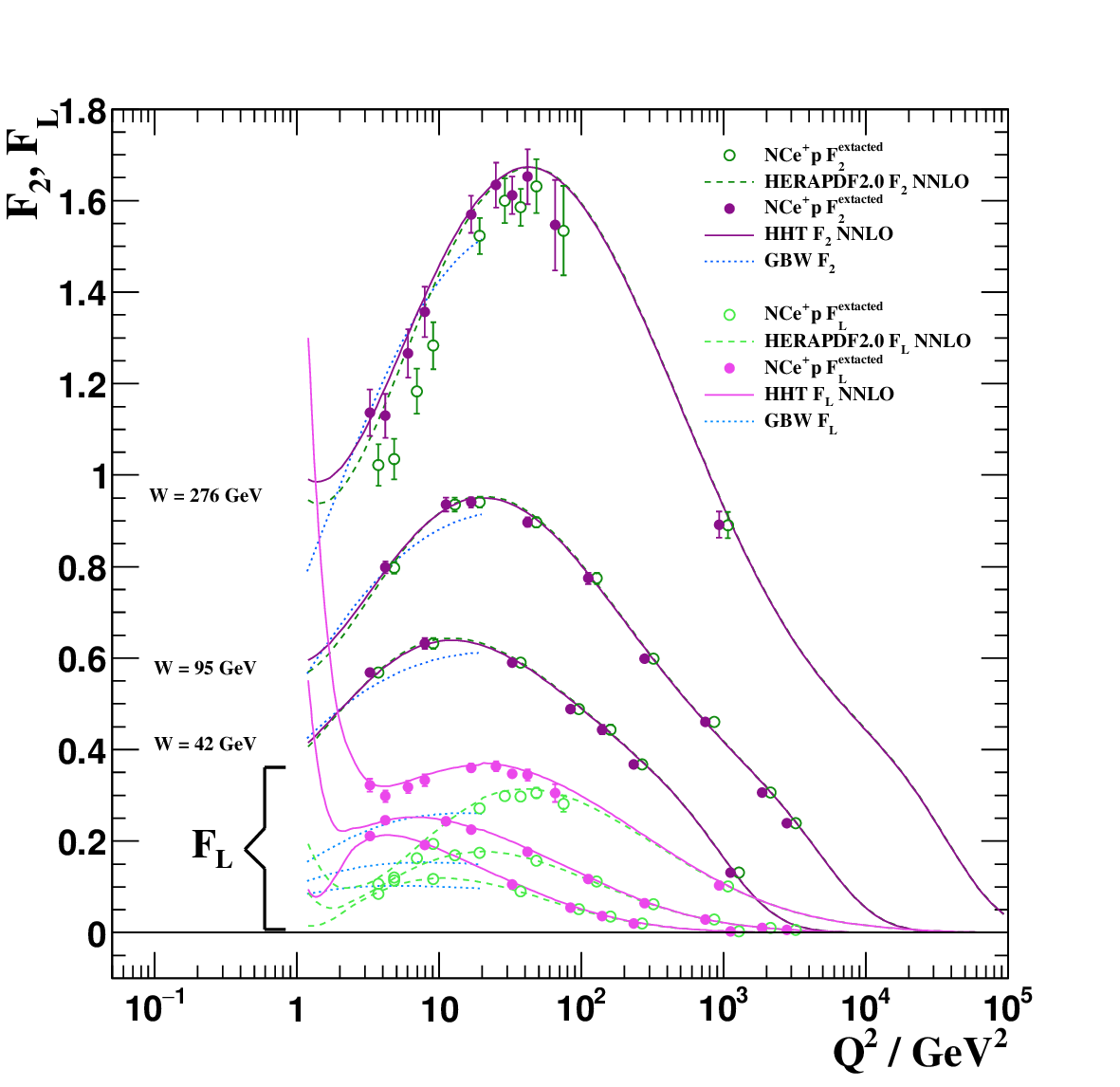  ,width=0.8\textwidth}}
\vspace{0.5cm}
\caption {The $F_2^{\rm extracted}$ and $F_{\rm L}^{\rm extracted}$ 
          values as extracted from
          HHT NNLO and HERAPDF2.0 NNLO together with 
          the corresponding predictions from HHT NNLO and HERAPDF2.0 NNLO  
          for three selected values of $W$.
          Also shown are predictions from the GBW model.
}
\label{fig:gb1}
\end{figure}

\begin{figure}[tbp]
\vspace{-0.3cm} 
\centerline{
\epsfig{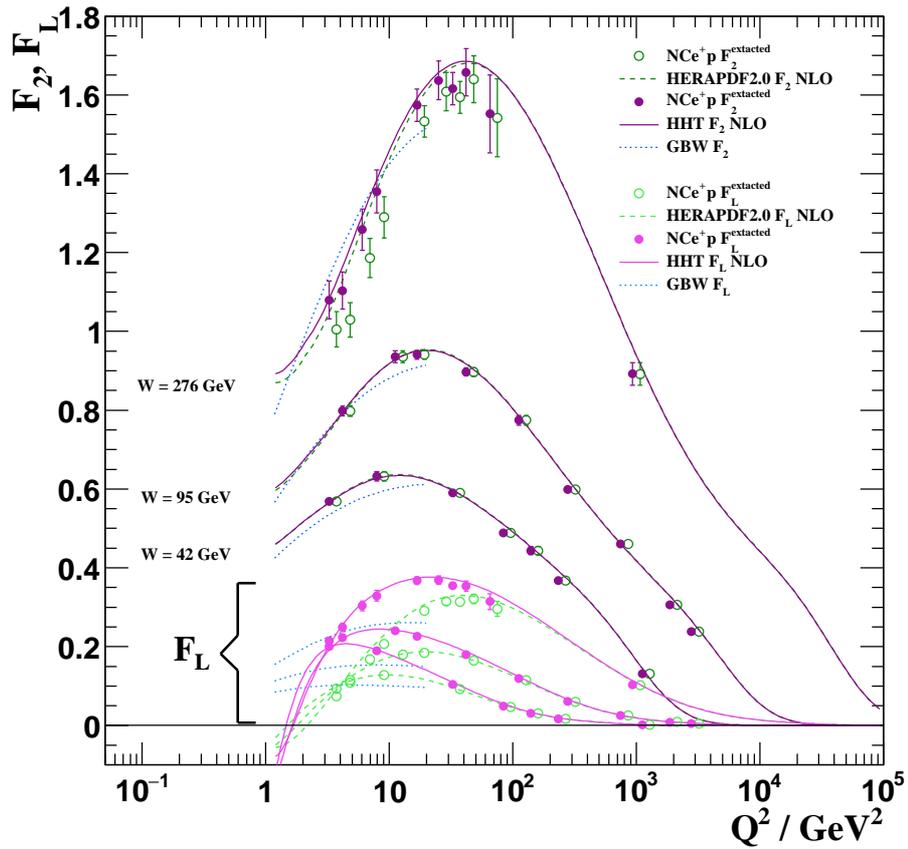}}
\vspace{0.5cm}
\caption {The $F_2^{\rm extracted}$ and $F_{\rm L}^{\rm extracted}$ 
          values as extracted from
          HHT NLO and HERAPDF2.0 NLO together with 
          the corresponding predictions from HHT NLO and HERAPDF2.0 NLO  
          for three selected values of $W$.
          Also shown are predictions from the GBW model.
}
\label{fig:gb2}
\end{figure}

Results on $F_2$ and $F_{\rm L}$ are presented for
selected values of
the energy at the photon--proton vertex, $W$,
to separate out the low-$x_{\rm Bj}$
regime of the data ($x_{\rm Bj} = Q^2/(W^2+Q^2)$) 
and to compare to the predictions of GBW.
Figures~\ref{fig:gb1} and~\ref{fig:gb2} show
extractions\,\footnote{Extracted values 
$F_{\rm L}^{\rm extracted}$ are calculated
similarly to the values of $F_2^{\rm extracted}$, see Eq.~\ref{eq:f2extr}.}
together with the corresponding predictions
for $F_2$ and $F_{\rm L}$
for the high-precision NC  $e^+p$ data
for HHT and HERAPDF2.0 at NNLO and NLO, respectively.
The data used here are limited to
$Q^2 \ge Q^2_{\rm min} = 3.5\,$GeV$^2$ and
approach the critical regime of $x_{\rm Bj}$ only for
$W=276$\,GeV.
The predictions of GBW, also shown in 
Figs.~\ref{fig:gb1} and~\ref{fig:gb2}, agree reasonably well
with the $F_2$ predictions of HHT up to $Q^2$ of about 10\,GeV$^2$
at this highest $W$ value, the only $W$ value
where HHT and HERAPDF2.0 differ
significantly. 
The values of $F_2^{\rm extracted}$ are significantly larger for 
HHT in this low-$x_{\rm Bj}$ regime than for HERAPDF2.0 and they agree
better with the corresponding predictions. This is true 
for fits at NNLO and at NLO.
In both cases, it is caused by significantly
larger values of $F_{\rm L}$, since Eq.~\ref{ncsi} implies that $F_2$
must also increase.

For Figs.~\ref{fig:gb1} and~\ref{fig:gb2},
all predictions from HHT and HERAPDF2.0 were extrapolated
down to $Q^2=1.2$\,GeV$^2$, a value
below the starting scale, for which the predictions of HHT NNLO
nevertheless still describe the reduced cross sections quite well,  
see Fig.~\ref{fig:sig_lowq2:1.2}.
The predictions from GBW are expected to be particularly relevant
in this regime while the pQCD evolution on which HHT and HERAPDF2.0
are based is expected to start to break down. 
This is demonstrated by the results on
$F_{\rm L}$.
The extractions and predictions differ substantially 
between NNLO and NLO for $Q^2$ below 10\,GeV$^2$.
At NLO, the predicted $F_{\rm L}$ values become negative 
for all three $W$ values as $Q^2$ approaches
1\,GeV$^2$ for both HERAPDF2.0 and HHT. 
This is unphysical.
At NNLO, all predicted $F_{\rm L}$ values start to increase as
$Q^2$ approaches 1\,GeV$^2$.
For HHT NNLO, this increase is dramatic.

Figures~\ref{fig:gb1} and~\ref{fig:gb2} also demonstrate
that values of $F_{\rm L}^{\rm extracted}$ cannot be considered
measurements.
Even though the predictions of HHT and HERAPDF2.0 differ
significantly below 100\,GeV$^2$, the extractions 
seem to simply reflect those predictions.
This demonstrates the importance of direct $F_{\rm L}$
measurements.

\begin{figure}[tbp]
\vspace{-0.3cm} 
\centerline{
\epsfig{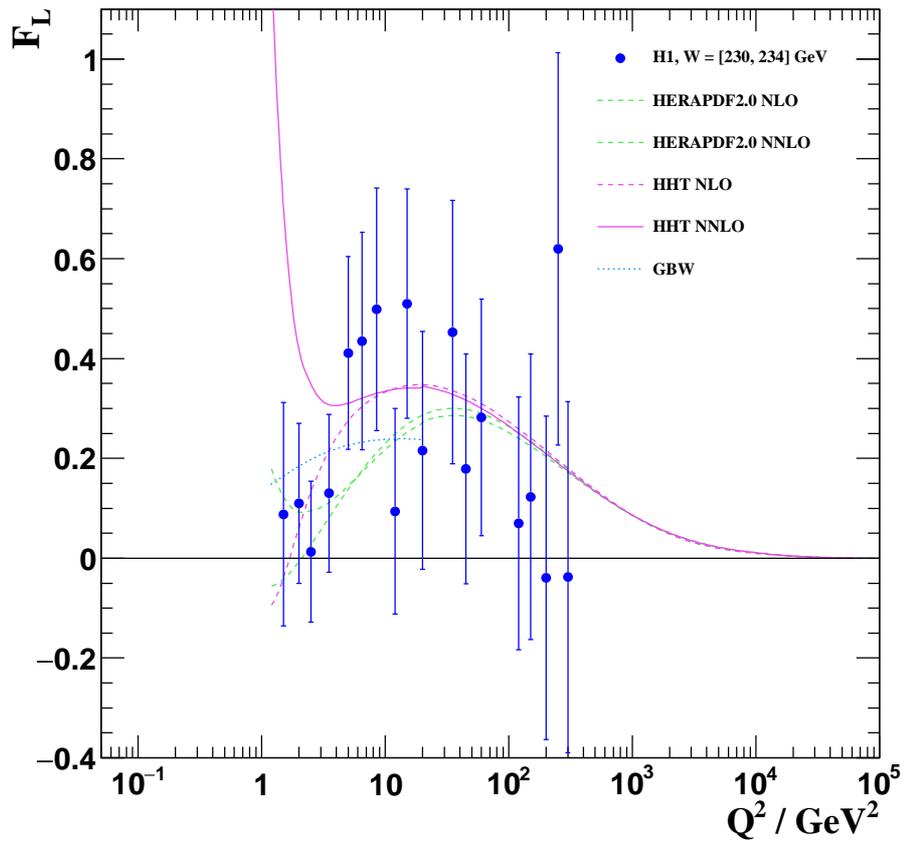}}
\vspace{0.5cm}
\caption {The predictions from HHT and HERAPDF2.0 at NNLO and NLO
          for $W=232\,$GeV, 
          together with direct measurements
          of $F_{\rm L}$ published by the H1 collaboration
          for $W$ between 230 and 234\,GeV.
          Also shown is the prediction from the GBW model.
}
\label{fig:gb3}
\end{figure}

Figure~\ref{fig:gb3} shows predictions for $F_{\rm L}$
from HHT and HERAPDF2.0 
at both NNLO and NLO for $W=232\,$GeV
together with  
a prediction from GBW.
The plot also contains
measured values 
down to $Q^2$ of almost 1\,GeV$^2$
published by the H1 collaboration~\cite{H1FL1}.
The statistical accuracy of these
data is limited, but the strong upturn of $F_{\rm L}$
predicted by HHT NNLO is not observed.
The data confirm
the downward trend of the $F_{\rm L}$ values measured for the full
$W$ range shown in Fig.~\ref{fig:flht}. 
Colour-dipole motivated models~\cite{gb,RT:FL:06} predict that $F_{\rm L}$
becomes similar for different values of $W$ at low $Q^2$.
The measured values of $F_{\rm L}$ shown in
Figs.~\ref{fig:flht} and~\ref{fig:gb3} are compatible with this.   

The strong difference between $F_{\rm L}$ predictions from HHT NNLO and NLO,
together with the HHT NNLO  prediction of a strong upturn of $F_{\rm L}$ as 
$Q^2$ approaches 1\,GeV$^2$ indicate that the current simple
higher-twist correction to the perturbative DGLAP
evolution alone cannot completely describe the physics involved, even though
the reduced cross sections are described quite well by this $ansatz$.

\section{Conclusions \label{sec:sum}}

The addition of a twist-4 term to the description of the
longitudinal structure function $F_{\rm L}$ significantly improved 
the quality of pQCD
fits within the DGLAP framework to HERA data.
In particular, the description of cross sections
at low $Q^2$ and low $x_{\rm Bj}$
was improved.
The $Q^2$ range of the fits was extended down to $Q^2=2.0$\,GeV$^2$
and the cross-section data could be well described 
down to $Q^2=1.2$\,GeV$^2$
by extrapolations.
The addition of a higher-twist term to the structure function
$F_2$ has no effect. This confirms the expectation that the influence of
higher-twist effects cancels for longitudinally and transversely
polarised photons in $F_2$.

The recombination of gluons is part of the higher-twist formalism.
This can be seen as a mechanism of saturation.
The strong influence of such a higher-twist term can be seen as
the first hint for the onset of saturation in the HERA data at 
low $Q^2$ and low $x_{\rm Bj}$.
The predictions of HHT NNLO for $F_{\rm L}$ become very high 
for $Q^2$ below 3.5\,GeV$^2$ and disagree with the data.
This indicates that the pQCD description is breaking down
and further mechanisms are needed for a consistent
description of the data at the lowest $x_{\rm Bj}$ and $Q^2$.

\section{Acknowledgements}

We are grateful to our ZEUS colleagues who supported
this work. We thank our funding agencies, especially the
Alexander von Humboldt foundation, 
for financial support and DESY for the hospitality extended 
to the non-DESY authors.
We also thank K.~Golec-Biernat for discussions and providing the
predictions of the GBW model.

\clearpage
\bibliography{HHT}
  
\clearpage
%

\begin{table}
\renewcommand*{\arraystretch}{1.2}
\begin{center}
  \begin{tabular}{llccc}
 \hline
 \hline
  Fit at &  with $Q^2_{\rm min}=3.5$\,GeV$^2$  &   HERAPDF2.0  &  HHT & $A_{\rm L}^{\rm HT}/$GeV$^2$ \\
\hline
NNLO & $\chi^2$/ndof      &   $1363/1131$         &  $1316/1130$   &  5.5$\pm$0.6 \\
     & $\chi^2$/ndp for NC $e^+p$:  $Q^2 \ge Q^2_{\rm min}$  
     &  $451/377$  & $422/377$  & \\
     & $\chi^2$/ndp for NC $e^+p$: $2.0$\,GeV$^2 \le Q^2 < Q^2_{\rm min}$ 
     & $41/25$ & $32/25$ & \\
\hline
 NLO & $\chi^2$/ndof       &  $1356/1131$          &  $1329/1130$ & 4.2$\pm$0.7 \\
     & $\chi^2$/ndp for NC $e^+p$: $Q^2 \ge Q^2_{\rm min}$   
     &  $447/377$    & $431/377$\\
     & $\chi^2$/ndp for NC $e^+p$: $2.0$\,GeV$^2 \le Q^2 < Q^2_{\rm min}$
     & $46/25$ & $46/25$ & \\
 \hline
 \hline
\end{tabular}
\end{center}
\caption{Table of $\chi^2$ values for the HHT fit compared to the
equivalent HERAPDF2.0 fit, both with $Q^2_{\rm min} = 3.5$\,GeV$^2$. 
Also listed are the partial $\chi^2$/(number of data points, ndp) 
 values of the fits 
for the high-precision
NC $e^+p$ data at $\sqrt{s}=318$\,GeV for $Q^2 \ge Q^2_{\rm min}$. 
The final row for each fit lists the $\chi^2$/ndp
for its predictions for $Q^2$ below the fitted region down to 2.0\,GeV$^2$.
In addition, the higher-twist parameters for HHT fits are given.
}
\label{tab:chitab}
\end{table}

\begin{table}
\renewcommand*{\arraystretch}{1.2}
\begin{center}
  \begin{tabular}{llccc}
 \hline
 \hline
  Fit at &  with $Q^2_{\rm min}=2.0$\,GeV$^2$  &   HERAPDF2.0  &  HHT & $A_{\rm L}^{\rm HT}/$GeV$^2$ \\
\hline
  NNLO  & $\chi^2$/ndof    & 1437/1171           &  $1381/1170$ & 5.2$\pm$0.7\\
        &  $\chi^2$/ndp for NC $e^+p$: 
          $Q^2 \ge Q^2_{\rm min}$       
                           & 486/402             & $457/402$ & \\
        &  $\chi^2$/ndp NC $e^+p$: 
          $Q^2_{\rm min} \le Q^2 < 3.5$\,GeV$^2$ 
                           & 31/25               &  26/25  & \\
 \hline
 NLO & $\chi^2$/ndof                 & 1433/1171 &  $1398/1170$ & 4.0$\pm$0.6 \\
  &  $\chi^2$/ndp for NC $e^+p$: 
    $Q^2 \ge Q^2_{\rm min}$       
                                      & 487/402 & $466/402$ & \\
  &  $\chi^2$/ndp NC $e^+p$: 
    $Q^2_{\rm min} \le Q^2 < 3.5$\,GeV$^2$ 
                                      & 40/25 & $31/25$ & \\
 \hline
 \hline
\end{tabular}
\end{center}
\caption{Table of $\chi^2$ values for the HHT fit compared to the
equivalent HERAPDF2.0 fit, both with $Q^2_{\rm min} = 2.0$\,GeV$^2$. 
Also listed are the partial $\chi^2$/ndp values of the fits 
for the high-precision
NC $e^+p$ data at $\sqrt{s}=318$\,GeV for $Q^2 \ge Q^2_{\rm min}$. 
The final row for each fit lists the partial $\chi^2$/ndp 
of the fit for data points 
with $2.0 \le Q^2 < 3.5$\,GeV$^2$.
In addition, the higher-twist parameters for HHT fits are given.
}
\label{tab:chitab2}
\end{table}


\clearpage


\begin{figure}[tbp]
\vspace{-0.3cm} 
\centerline{
\epsfig{figure=figures/ncep-nlo1-ht-q2gt2.eps ,width=0.9\textwidth}}
\centerline{
\epsfig{figure=figures/ncep-nlo-herapdf2.0-q2gt2.eps,width=0.9\textwidth}}
\vspace{0.5cm}
\caption {The predictions of HHT NLO (top) and HERAPDF2.0 NLO (bottom), 
         both with $Q^2_{\rm min} = 3.5$\,GeV$^2$, 
         compared to the HERA measurements of $\sigma_{r}$. 
         The bands represent the experimental, i.e.\ fit, uncertainties.
         Extrapolations are indicated as dotted lines.
}
\label{fig:NLO:HHT-HPDF}
\end{figure}


\begin{figure}[tbp]
\vspace{-0.3cm} 
\centerline{
\epsfig{figure=figures/f2-nlo1-ht-fitq2gt3.5.eps ,width=0.7\textwidth}}
\centerline{
\epsfig{figure=figures/f2-nlo-herapdf2.0-fitq2gt3.5.eps  ,width=0.7\textwidth}}
\vspace{0.5cm}
\caption {The predictions of HHT NLO (top) and HERAPDF2.0 NLO (bottom) 
         both with $Q^2_{\rm min} = 3.5$\,GeV$^2$, 
         compared to extracted values  $F_2^{\rm extracted}$. 
         The bands represent the experimental, i.e.\ fit, uncertainties.
         Extrapolations are indicated as dotted lines.
}
\label{fig:f2:NLO:HHT-HPDF}
\end{figure}


\clearpage

\clearpage

\end{document}